\documentstyle[prl,aps,twocolumn]{revtex}

\begin{document} 
\draft 
\title{Semiclassical Calculation of Transition Matrix Elements 
       for Atoms in External Fields}
\author{J\"org Main$^1$ and G\"unter Wunner$^2$}
\address{$^1$Institut f\"ur Theoretische Physik I,
         Ruhr-Universit\"at Bochum, D-44780 Bochum, Germany}
\address{$^2$Institut f\"ur Theoretische Physik und Synergetik,
         Universit\"at Stuttgart, D-70550 Stuttgart, Germany}
%
\date{\today}
\maketitle

\begin{abstract}
Closed orbit theory is generalized to the semiclassical calculation of
cross-correlated recurrence functions for atoms in external fields.
The cross-correlation functions are inverted by a high resolution spectral
analyzer to obtain the semiclassical eigenenergies and transition matrix
elements.  The method is demonstrated for dipole transitions of the 
hydrogen atom in a magnetic field.  This is the first semiclassical 
calculation of individual quantum transition strengths from closed orbit 
theory.
\end{abstract}


In 1925, Heisenberg \cite{Hei25} pushed open the door to quantum mechanics 
when he proposed a quantum theory consisting only of 
``in principle observable'' quantities -- matrix elements $A(n,m)$ --, 
whose physical meaning was hinged upon the correspondence principle:
For high quantum numbers $n$ and  $\vert n-m \vert \ll n$ the matrix elements
were  to turn into the Fourier amplitudes of the corresponding 
physical observable $A(t)$  associated with a classical periodic
orbit, viz.\
$A(n,m)\exp(i(E(n)-E(m))t/\hbar) \Leftrightarrow 
A_\tau(n) \exp(i\tau \omega(n))$, with $\tau = n-m$ and
$\omega(n)$ the frequency of the periodic orbit. 
By a consistent application of the translation rules to the  
quantization condition for actions, $\Delta S/\Delta n = h$, he arrived
at the canonical commutation relations, underlying the whole of quantum 
mechanics. Thus establishing the connection between the information 
encoded in classical orbits and quantum mechanical transitions 
amplitudes proved one of the fundamental questions of quantum physics. 

The correspondence principle is silent about transition amplitudes
involving low-lying states. It was, therefore, a great success when,
more than sixty years later, {\em closed orbit theory} -- a variant of 
periodic orbit theory \cite{Gut90} --
came up with the discovery that there exists an intimate connection between 
classical orbits and transitions amplitudes even in cases where one of the 
states lies in the deep quantum regime. In closed orbit theory, developed    
by Du and Delos \cite{Du87} and Bogomolny \cite{Bog88}, the transition
amplitudes  are given as the sum of two terms, one a smoothly varying part 
(as a function energy) and the other a superposition of sinusoidal modulations.
The frequencies, amplitudes, and phases of the modulations are directly
obtained from information contained in the closed classical orbits.
When the resulting transition amplitudes are Fourier transformed the 
sinusoidal modulations produce sharp peaks in the Fourier transform 
recurrence spectra. 
Closed orbit theory has been applied to the interpretation of photoabsorption 
spectra of atoms in external fields and has been most successful in explaining
the quantum mechanical recurrence spectra qualitatively and even 
quantitatively in terms of the closed orbits of the underlying classical 
system \cite{Mai91,Vel93,Cou94}.

However, up to now practical applications of closed orbit theory have always 
been restricted to the semiclassical calculation of {\em low resolution} 
spectra for two reasons.
Firstly, the closed orbit sum requires, in principle, the knowledge of all
orbits up to infinite length, which are not normally available from a
numerical closed orbit search, and, secondly, the infinite closed orbit sum 
suffers from fundamental convergence problems \cite{Du87,Bog88}.
It is therefore usually believed that the calculation of {\em individual} 
transition matrix elements, e.g., of the dipole 
operator $D$, $\langle\phi_i|D|\psi_f\rangle$, which describe the transition 
strengths from the initial state 
$|\phi_i\rangle$ to final states $|\psi_f\rangle$, is a problem beyond
the applicability of the semiclassical closed orbit theory, i.e., is the 
domain of quantum mechanical methods.

It is the purpose of this Paper to demonstrate that 
high-precision quantum transition amplitudes between low-lying and highly
excited states can be obtained within the framework of closed orbit
theory using solely the information contained in closed classical orbits.
In this way we can establish a connection between classical orbits and 
quantum mechanical matrix elements that goes far beyond what was
conceived of in the early days of quantum mechanics.  
To that end, we  slightly  generalize closed orbit theory to the semiclassical
calculation of cross-correlated recurrence functions.
We then adopt the method of Refs.\ \cite{Wal95,Nar97,Man98a} to harmonically
invert the cross-correlated recurrence signal and to extract the semiclassical
eigenenergies and transition matrix elements.
Results will be presented for the photo excitation of the hydrogen atom in
a magnetic field.

The oscillator strength $f$ for the photo excitation of atoms in external 
fields can be written as
\begin{equation}
 f(E) = -{2\over\pi} (E-E_i) \:
   {\rm Im}\, \langle\phi_i|D G_E^+ D|\phi_i\rangle \; ,
\label{fE:eq}
\end{equation}
where $|\phi_i\rangle$ is the initial state at energy $E_i$, $D$ is the
dipole operator, and $G_E^+$ the retarded Green's function of the atomic 
system.
The basic steps for the derivation of closed orbit theory are to replace
the quantum mechanical Green's function in (\ref{fE:eq}) with its
semiclassical Van Vleck-Gutzwiller approximation and to carry out the
overlap integrals with the initial state $|\phi_i\rangle$.
Here we go one step further by introducing a cross-correlation matrix
\begin{equation}
 g_{\alpha\alpha'} = \langle\phi_\alpha|DG_E^+D|\phi_{\alpha'}\rangle
\label{g_def:eq}
\end{equation}
with $|\phi_\alpha\rangle$, $\alpha=1,2,\dots,L$ a set of independent initial 
states.
As will be shown below the use of cross-correlation matrices can considerably
improve the convergence properties of the semiclassical procedure.
In the following we will concentrate on the hydrogen atom in a magnetic field
(for reviews see \cite{Fri89,Has89,Wat93})
with $\gamma=B/(2.35 \times 10^5\,{\rm T})$ the magnetic field strength
in atomic units.
The system has a scaling property, i.e., the shape of periodic orbits does
not depend on the scaling parameter, $w=\gamma^{-1/3}=\hbar_{\rm eff}^{-1}$,
and the classical action scales as $S=sw$, with $s$ the scaled action.
As, e.g.,\ in Ref.\ \cite{Mai91}, we consider scaled photoabsorption spectra 
at constant scaled energy $\tilde E=E\gamma^{-2/3}$ as a function of the 
scaling parameter $w$.
We choose dipole transitions between states with magnetic quantum number
$m=0$.
Note that the following ideas can be applied in an analogous way to atoms
in electric fields. 
Following the derivation of Refs.\ \cite{Du87,Bog88} the semiclassical
approximation to the fluctuating part of $g_{\alpha\alpha'}$ in Eq.\
\ref{g_def:eq} reads
\begin{eqnarray}
     g_{\alpha\alpha'}^{\rm sc}(w)
 &=& w^{-1/2} \sum_{{\rm co}} {-(2\pi)^{5/2}\over\sqrt{|m_{12}^{\rm co}|}}
    \sqrt{\sin\vartheta_i^{\rm co}\sin\vartheta_f^{\rm co}} \nonumber \\
 &\times&
    {\cal Y}_\alpha(\vartheta_i^{\rm co})
    {\cal Y}_{\alpha'}(\vartheta_f^{\rm co}) \,
    e^{i\left(s_{\rm co}w-{\pi\over 2}\mu_{\rm co}+{\pi\over 4}\right)} \; ,
\label{g_sc:eq}
\end{eqnarray}
with $s_{\rm co}$ and $\mu_{\rm co}$ the scaled action and Maslov index
of the closed orbit (co), $m_{12}^{\rm co}$ an element of the monodromy
matrix, and $\vartheta_i^{\rm co}$ and $\vartheta_f^{\rm co}$ the initial
and final angle of the trajectory with respect to the magnetic field axis.
The angular functions ${\cal Y}_\alpha(\vartheta)$ depend on the states
$|\phi_\alpha\rangle$ and the dipole operator $D$, and are given as a
linear superposition of Legendre polynomials,
${\cal Y}_\alpha(\vartheta)=\sum_l{\cal B}_{l\alpha} P_l(\cos\vartheta)$.
For low-lying initial states with principal quantum number $n$ only few 
coefficients ${\cal B}_{l\alpha}$ with $l\le n$ are nonzero.
Explicit formulas for the calculation of the coefficients can be found in
Refs.\ \cite{Du87,Bog88}.
The problem now is to extract the semiclassical eigenenergies and transition
matrix elements from Eq.\ \ref{g_sc:eq} because the closed orbit sum does
not converge.
We therefore adopt the idea of Ref.\ \cite{Mai97} where we proposed 
to adjust the Fourier transform of a non-convergent Dirichlet series
like the semiclassical expression (\ref{g_sc:eq}) to the functional form
of its quantum mechanical analogue.
The Fourier transformation of $w^{1/2}g_{\alpha\alpha'}^{\rm sc}(w)$
yields the cross-correlated recurrence signals
\begin{equation}
   C_{\alpha\alpha'}^{\rm sc}(s)
 = \sum_{{\rm co}} {\cal A}_{\alpha\alpha'}^{\rm co} \delta(s - s_{\rm co})\; ,
\label{C_sc:eq}
\end{equation}
with the amplitudes
\begin{eqnarray}
     {\cal A}_{\alpha\alpha'}^{\rm co}
 &=& {-(2\pi)^{5/2}\over\sqrt{|m_{12}^{\rm co}|}}
      \sqrt{\sin\vartheta_i^{\rm co}\sin\vartheta_f^{\rm co}}
  \nonumber \\ &\times&
      {\cal Y}_{\alpha }(\vartheta_i^{\rm co})
      {\cal Y}_{\alpha'}(\vartheta_f^{\rm co}) \,
      e^{i\left(-{\pi\over 2}\mu_{\rm co}+{\pi\over 4}\right)}
\label{ampl:eq}
\end{eqnarray}
being determined exclusively by closed orbit quantities.
The corresponding quantum mechanical cross-correlated recurrence functions, 
i.e., the Fourier transforms of $w^{1/2}g_{\alpha\alpha'}^{\rm qm}(w)$ read
\begin{equation}
   C^{\rm qm}_{\alpha\alpha'}(s)
 = -i \sum_k b_{\alpha k} b_{\alpha' k} \, e^{-iw_ks} \; ,
\label{C_qm:eq}
\end{equation}
with $w_k$ the eigenvalues of the scaling parameter, and
\begin{equation}
 b_{\alpha k} = w_k^{1/4} \langle\phi_\alpha|D|\psi_k\rangle
\end{equation}
proportional to the transition matrix element for the transition from the 
initial state $|\phi_\alpha\rangle$ to the final state $|\psi_k\rangle$.

The method to adjust (\ref{C_sc:eq}) to (\ref{C_qm:eq}) for fixed states
$|\phi_\alpha\rangle$ and $|\phi_{\alpha'}\rangle$ is that of 
harmonic inversion as discussed in Ref.\ \cite{Mai97}.
However, information theoretical considerations then yield an estimate for
the required signal length, $s_{\rm max}\sim 4\pi\bar\varrho(w)$ \cite{Mai97},
which may result in an unfavorable scaling because of a rapid proliferation
of closed orbits with increasing period.
Moreover, it is a  special problem to resolve nearly degenerate states and to
detect states with very low transition strengths from the harmonic inversion
of a {\em single} function $C^{\rm sc}_{\alpha\alpha'}(s)$.
Note also that the element of the monodromy matrix $m_{12}$ and the
values of the angular functions ${\cal Y}_\alpha(\vartheta_i)$ and 
${\cal Y}_{\alpha'}(\vartheta_f)$ are intrinsically intertwined in 
Eq.\ \ref{ampl:eq} for the amplitudes ${\cal A}_{\alpha\alpha'}^{\rm co}$,
i.e., a {single} function $C^{\rm sc}_{\alpha\alpha'}(s)$ does not contain
independently the information from the monodromy matrix and the starting
and returning angles $\vartheta_i$ and $\vartheta_f$ of the closed orbits.

In this Paper we therefore propose to apply an extension of the method to the
harmonic inversion of cross-correlation functions \cite{Wal95,Nar97,Man98a},
which has recently also served as a powerful tool for the semiclassical
calculation of tunneling splittings \cite{Man98b}.
The idea is that the informational content of an $L \times L$ time signal
is increased roughly by a factor of $L$ as compared to a $1 \times 1$ signal.
The additional information is gained with the set of linearly independent
angular functions ${\cal Y}_\alpha(\vartheta)$, $\alpha=1,2,\dots L$ in
Eq.\ \ref{ampl:eq} evaluated at the starting and returning angles 
$\vartheta_i$ and $\vartheta_f$ of the closed orbits.
Note that the cross-correlation matrix (\ref{C_sc:eq}) is constructed by
using independently the information of the closed orbit quantities, i.e.,
the elements $m_{12}$ of the monodromy matrix and the angles $\vartheta_i$ 
and $\vartheta_f$.
For a given number of closed orbits the accuracy of semiclassical 
spectra can be significantly improved with the help of the cross-correlation 
approach, or, alternatively, spectra with similar accuracy can be obtained 
from a closed orbit cross-correlation signal with a significantly reduced 
signal length.

Here we only give a qualitative and brief description of the method.
The details of the numerical procedure of solving the generalized 
harmonic inversion problem (\ref{C_qm:eq}) have been presented in 
Refs.\ \cite{Wal95,Nar97,Man98a}.
The idea is to recast the nonlinear fit problem as a linear algebraic
problem \cite{Wal95}. 
This is accomplished by associating the signal $C_{\alpha\alpha'}(s)$
\newpage
\phantom{}
\begin{figure}
\vspace{9.7cm}
\includegraphics{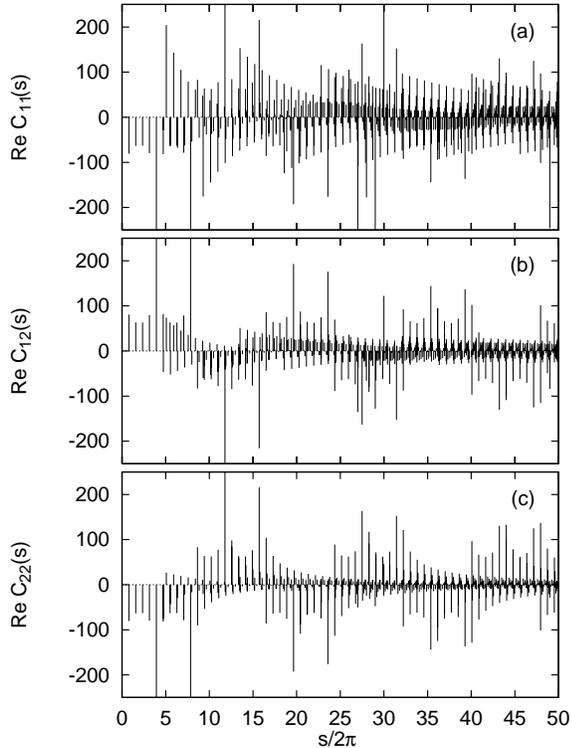}
\caption{\label{fig1} 
Real parts of the semiclassical cross-correlated recurrence functions for the 
hydrogen atom in a magnetic field at constant scaled energy $\tilde E=-0.7$
as functions of the classical action $s$ (in dimensionless scaled units).
The graphs of the imaginary parts (not shown) qualitatively resemble those 
of the real parts.
}
\end{figure}
\noindent
(to be inverted) with a time cross-correlation function between an initial 
state $\Phi_{\alpha}$ and a final state $\Phi_{\alpha'}$,
\begin{equation}
   C_{\alpha\alpha'}(s)
 = \langle\Phi_{\alpha'}|e^{-is\hat H_{\rm eff}}\Phi_{\alpha}\rangle \; ,
\label{ansatz}
\end{equation}
where the fictitious quantum dynamical system is described by an effective 
Hamiltonian $\hat H_{\rm eff}$.
The latter is defined implicitly by relating its spectrum to the set of 
unknown spectral parameters $w_k$ and $b_{\alpha k}$.
Diagonalization of $\hat H_{\rm eff}$ would yield the desired $w_k$ and 
$b_{\alpha k}$. 
This is done by introducing an appropriate basis set in which the matrix 
elements of $\hat H_{\rm eff}$ are available only in terms of the known
signals $C_{\alpha\alpha'}(s)$.
The Hamiltonian $\hat H_{\rm eff}$ is assumed to be complex symmetric even
in the case of a bound system, which makes the harmonic inversion stable 
with respect to ``noise'' due to the imperfections of the semiclassical 
approximation.

We now demonstrate the method of harmonic inversion of the cross-correlated
closed orbit recurrence functions (\ref{C_sc:eq}) for the example of the
hydrogen atom in a magnetic field at constant scaled energy $\tilde E=-0.7$.
This energy was also chosen for detailed experimental investigations of
the helium atom \cite{Vel93}.
We investigate dipole transitions from the initial state 
$|\phi_1\rangle = |2p0\rangle$ with light polarized parallel to the 
magnetic field axis to final states with 
\newpage
\phantom{}
\begin{figure}
\vspace{9.8cm}
\includegraphics{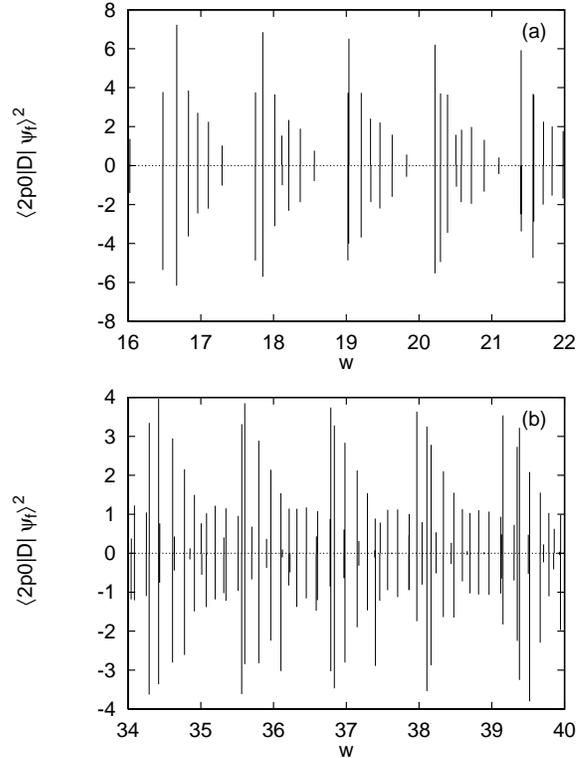}
\caption{\label{fig2} 
Quantum (upper part) and semiclassical (lower part) photoabsorption spectra
of the hydrogen atom in a magnetic field at scaled energy $\tilde E=-0.7$
as functions of the dimensionless scaling parameter $w=\gamma^{-1/3}$:
Transition matrix elements $\langle 2p0|D|\psi_f\rangle^2$ for dipole
transitions with light polarized parallel to the magnetic field axis.
}
\end{figure}
\noindent
magnetic quantum number $m=0$.
For this transition the angular function in Eq.\ \ref{ampl:eq} reads
${\cal Y}_1(\vartheta)=(2\pi)^{-1/2}2^7e^{-4}(4\cos^2\vartheta-1)$.
For the construction of a $2 \times 2$ cross-correlated recurrence signal
we use for simplicity as a second transition formally an outgoing $s$-wave, 
i.e., $D|\phi_2\rangle \propto Y_{0,0}$, and, thus, 
${\cal Y}_2(\vartheta)=\mbox{const}$.
A numerical closed orbit search yields 1395 primitive closed orbits 
(2397 orbits including repetitions) with scaled action $s/2\pi<100$.
With the closed orbit quantities at hand it is straightforward to calculate
the cross-correlated recurrence functions in (\ref{C_sc:eq}).
The real parts of the functions $C_{11}^{\rm sc}(s)$, 
$C_{12}^{\rm sc}(s)$, and $C_{22}^{\rm sc}(s)$ with $s/2\pi<50$ are 
presented in Fig.\ \ref{fig1}.
The imaginary parts are not shown because they qualitatively resemble the 
real parts.
Note that for symmetry reasons $C_{21}^{\rm sc}(s)=C_{12}^{\rm sc}(s)$.

We have inverted the $2 \times 2$ cross-correlated recurrence functions
in the region $0<s/2\pi<100$.
The resulting semiclassical photoabsorption spectrum is compared 
with the exact quantum spectrum in Fig.\ \ref{fig2}a for the region $16<w<21$
and in Fig.\ \ref{fig2}b for the region $34<w<40$.
The upper and lower parts in Fig.\ \ref{fig2} show the exact quantum spectrum 
and the semiclassical spectrum, respectively.
Note that the region of the spectrum presented in Fig.\ \ref{fig2}b belongs
well to the experimentally accessible regime with laboratory field strengths
$B=6.0 \, {\rm T}$ to $B=3.7 \,{\rm T}$.
The overall agreement between 
the quantum and semiclassical spectrum is impressive, even though 
a line by line comparison still reveals small differences for a few  
matrix elements. 
It is important to note that the high quality of the semiclassical spectrum
could only be achieved by our application of the cross-correlation approach.
For example, the two nearly degenerate states at $w=36.969$ and $w=36.982$
cannot be resolved and the very weak transition at $w=38.894$ with
$\langle2p0|D|\psi_f\rangle^2=0.028$ is not detected with a single 
$(1 \times 1)$ recurrence signal of the same length.
However, these hardly visible details are indeed present in the semiclassical 
spectrum in Fig.\ \ref{fig2}b obtained from the harmonic inversion of the 
$2 \times 2$ cross-correlated recurrence functions.

In conclusion, we have demonstrated that closed orbit theory is not 
restricted to describe long-range  modulations in quantum mechanical
photoabsorption spectra of atoms in external fields but can well be 
applied to extract individual eigenenergies and transition matrix 
elements from the closed orbit quantities.
This is achieved by a high resolution spectral analysis (harmonic inversion)
of cross-correlated closed orbit recurrence signals.
For the hydrogen atom in a magnetic field we have obtained, for the first
time, transition matrix elements between low-lying and highly excited
Rydberg states using exclusively  classical closed orbit data.
It will be straightforward, and rewarding, to apply the method to atoms
in electric fields.

\bigskip
We acknowledge fruitful discussions with V.\ Mandelshtam.
This work was supported in part by the Son\-der\-for\-schungs\-be\-reich 
No.\ 237 of the Deutsche For\-schungs\-ge\-mein\-schaft.
J.M.\ thanks the Deutsche For\-schungs\-ge\-mein\-schaft for a 
Habilitandenstipendium (Grant No.\ Ma 1639/3).


\begin{references}
\bibitem{Hei25}  W. Heisenberg, Z. Physik {\bf 33}, 879 (1925);
                 M. Born,  W. Heisenberg, P. Jordan, Z. Physik {\bf 35},
                 557 (1926).
\bibitem{Gut90}  M. C. Gutzwiller, {\it Chaos in Classical and Quantum
                 Mechanics} (Springer, New York, 1990).
\bibitem{Du87}   M. L. Du and J. B. Delos, Phys. Rev. Lett. {\bf 58}, 1731
                 (1987) and Phys. Rev. A {\bf 38}, 1896 and 1913 (1988).
\bibitem{Bog88}  E. B. Bogomolny, JETP Lett. {\bf 47}, 526 (1988) and
                 Sov. Phys. JETP {\bf 69}, 275 (1989).
\bibitem{Mai91}  J. Main, G. Wiebusch, and K. H. Welge, Comm. At. Mol. Phys.
                 {\bf 25}, 233 (1991);
                 J. Main, G. Wiebusch, K. H. Welge, J. Shaw, and J. B. Delos,
                 Phys. Rev. A {\bf 49}, 847 (1994).
\bibitem{Vel93}  T. van der Veldt, W. Vassen, and W. Hogervorst,
                 Europhys. Lett. {\bf 21}, 903 (1993).
\bibitem{Cou94}  M. Courtney, H. Jiao, N. Spellmeyer, and D. Kleppner,
                 Phys. Rev. Lett. {\bf 73}, 1340 (1994);
                 M. Courtney, N. Spellmeyer, H. Jiao, and D. Kleppner,
                 Phys. Rev. A {\bf 51}, 3604 (1995).
\bibitem{Wal95}  M. R. Wall and D. Neuhauser, J. Chem. Phys. {\bf 102},
                 8011 (1995).
\bibitem{Nar97}  E. Narevicius, D. Neuhauser, H. J. Korsch, and N. Moiseyev, 
                 Chem. Phys. Lett. {\bf 276}, 250 (1997).
\bibitem{Man98a} V. A. Mandelshtam, J. Chem. Phys. {\bf 108}, 9999 (1998).
\bibitem{Fri89}  H. Friedrich and D. Wintgen, Phys. Rep. {\bf 183}, 37 (1989).
\bibitem{Has89}  H. Hasegawa, M. Robnik, and G. Wunner, Prog. Theor. Phys.
                 Suppl. {\bf 98}, 198 (1989).
\bibitem{Wat93}  S. Watanabe, {\it Review of Fundamental Processes and
                 Applications of Atoms and Ions}, ed. C. D. Lin,
                 (World Scientific, Singapore, 1993).
\bibitem{Mai97}  J. Main, V. A. Mandelshtam, and H. S. Taylor, Phys. Rev.
                 Lett. {\bf 79}, 825 (1997);
                 J. Main, V. A. Mandelshtam, G. Wunner, and H. S. Taylor,
                 Nonlinearity {\bf 11}, 1015 (1998).
\bibitem{Man98b} V. A. Mandelshtam and M. Ovchinnikov, J. Chem. Phys.
                 {\bf 108}, 9206 (1998).
\end{references}
\end{document}